\documentclass[a4paper,11pt]{article}
\usepackage{pos}

\renewenvironment{thebibliography}[1]
{\begin{oldthebibliography}{#1}
 \setlength{\itemsep}{0pt}
 \setlength{\parskip}{0pt}
 \setlength{\parsep}{0pt}}
{\end{oldthebibliography}}

\usepackage{enumitem}
\setlist{nosep,leftmargin=*}

\usepackage{titlesec}
\titlespacing*{\section}{0pt}{0.8ex plus 0.2ex minus 0.1ex}{0.4ex}
\titlespacing*{\subsection}{0pt}{0.6ex plus 0.2ex minus 0.1ex}{0.3ex}

\title{Quantum Simulation of Gauge Theories for Particle and Nuclear Physics
}
\ShortTitle{Quantum Simulation of Gauge Theories for Particle and Nuclear Physics}

\author*[a,b]{Zohreh Davoudi}

\affiliation[a]{Maryland Center for Fundamental Physics and Department of Physics, University of Maryland, College Park, MD 20742, USA}

\affiliation[b]{Joint Center for Quantum Information and Computer Science, University of Maryland, College Park, Maryland 20742, USA}

\emailAdd{davoudi@umd.edu}

\abstract{Lattice field theory, along with its algorithmic and hardware ecosystems, has been at the forefront of computational particle and nuclear physics. It continues to deliver impressive results on the hadronic spectrum, structure, decays, and reactions. Yet, this vigorous campaign has fallen short in addressing a range of problems involving dense matter and general dynamical phenomena. The reason is that such problems require an exponential scaling of computing time and space in system size. Quantum simulation, enabled by quantum-computing algorithms and hardware technology, promises a way forward by offering several polynomially efficient algorithms compared with their inefficient classical counterparts. Lattice gauge theorists have engaged in a multi-pronged program to leverage such new possibilities, and have steadily advanced the state of theory, algorithm, and hardware implementations and co-design. In this talk, I motivate the quantum-computational lattice-field-theory program; introduce the questions such a program is expected to address and the strategies it involves; report on recent progress; and end with a note on challenges and opportunities ahead.}

\renewcommand{\hookAfterAbstract}{
  \par\bigskip
  \noindent\textsc{Preprint}: UMD-PP-026-03, INT-PUB-26-018
}

\FullConference{The 42nd International Symposium on Lattice Field Theory (LATTICE2025)\\
2-8 November 2025\\
Tata Institute of Fundamental Research, Mumbai, India\\}

\begin{document}
\maketitle

\section{Motivation and background}

An overarching goal of particle- and nuclear-physics research is to predict the complexities of the visible sector of the universe rooted in the fundamental quantum field theory of forces and particles, namely the Standard Model (SM). Examples of such complexities include reactions and structure of ordinary and rare atomic nuclei, exotic phases of strongly interacting matter, such as those conjectured to exist in the interior of neutron stars, and supernovae evolution and the origin of heavy elements. Another goal is to discover the physics beyond the SM in the quest to explain dark matter and dark energy, matter-antimatter asymmetry, and the nature of neutrinos---questions that often require isolating the minuscule effects of new physics in ordinary matter from those of the SM.

Our best tool to incorporate SM interactions, more precisely quantum chromodynamics (QCD) and quantum electrodynamics (QED), in studies of hadrons, nuclei, and thermal matter is lattice (gauge) field theory. The method relies on Monte-Carlo-sampling techniques, which allow nonperturbative computation of observables on a discretized finite Euclidean spacetime, starting from a Lagrangian or an action. Lattice field theory has led to an impressive array of verifications and predictions, from the spectrum and structure of hadrons and nuclei, few-hadron scattering, decay, and reaction amplitudes, and the equation of state of dilute matter (see Refs.~\cite{FlavourLatticeAveragingGroupFLAG:2024oxs,Davoudi:2022bnl} for select reviews).

Yet, lattice-field-theory techniques fall short when applied to several systems and phenomena: 
\begin{itemize}
\item[$\diamond$] \emph{Atomic nuclei with a sizable number of nucleons}: The complexity of nuclear correlation functions roughly scales factorially with the number of quark constituents; such correlation functions suffer an exponentially decaying signal as a function of the Euclidean time and the number of nucleons; and the excitation gaps decrease substantially as a function of atomic number, rendering the isolation of low-lying excitations challenging.
\item[$\diamond$] \emph{Finite-density phases of matter}: Due to a fermion sign problem in Monte-Carlo computations of systems at a finite baryon density, the phase diagram of strongly interacting matter remains largely under speculation. As a result, reliable predictions for the existence of exotic phases at high densities, the nature of phase transitions, and the existence and location of critical points remain limited.
\item[$\diamond$] \emph{Evolution and equilibration of matter}: Euclidean spacetime enables sign-problem-free Monte-Carlo computations (of dilute matter) but prevents access to real-(Minkowski-)time phenomena (except in limited cases~\cite{Luscher:1986pf}). As a result, it remains challenging to predict, using first-principles means, how phases of matter are produced, evolve, and ultimately convert to hadrons, in the early universe or after the collision of energetic hadrons in colliders. 
\item[$\diamond$] \emph{Dynamical structure and response functions}: Similarly, any observable whose definition inherently involves Minkowski time cannot be accessed directly via Euclidean Monte-Carlo computations (except in certain kinematic regimes~\cite{Ji:2013dva}). These include hadron tensor and various hadronic distribution functions, medium's transport coefficients, and dynamical response functions (of, e.g., atomic nuclei in neutrino experiments).
\item[$\diamond$] \emph{Entanglement and quantum correlations}: Entanglement structure of states, which ultimately informs, e.g., the nature of quantum gravity from a gauge-gravity duality perspective~\cite{Ryu:2006bv,VanRaamsdonk:2010pw}, topological phases of matter~\cite{Kitaev:2005dm,Levin:2006zz,Li:2008kda}, thermalization dynamics~\cite{Mueller:2021gxd} and thermodynamic quantities~\cite{Davoudi:2024osg}, is hard to discern in Euclidean Monte-Carlo simulations.
\end{itemize}
%

The rapid development of quantum-simulation and quantum-computing paradigms promises more efficient routes to addressing the problems above. The reason is twofold. On one hand, quantum Hilbert spaces can be encoded far more efficiently in quantum units (such as qubits). On the other hand, operations can be parallelized significantly by leveraging quantum superposition and entanglement. In fact, tracking the amplitudes of time-evolved quantum states  in real time is a natural task in quantum hardware: unitary dynamics in quantum systems can be leveraged to simulate time evolution with resources that scale polynomially in system size~\cite{Feynman:1982,Lloyd:1996}.

Theoretical promise is not sufficient for meaningful progress, and the state of hardware is an important factor in ensuring success. Luckily, quantum-hardware technology for simulation and computing purposes has witnessed a significant leap over the past decade~\cite{Altman:2019vbv}. The field has come a long way from first demonstrations of high-fidelity operations in a few-qubit systems, to circuit implementations of basic algorithms, to nontrivial simulations on tens of qubits, to first implementations of error-corrected qubits and circuits~\cite{Preskill:2018jim,Regent:2025cos}. Nowadays, trapped ions and superconducting circuits present the most promising architectures in digital, gate-based computing; cold atoms in optical lattices have shown success in analog quantum simulation of several many-body quantum systems; and Rydberg arrays are increasingly showing promise in both analog and digital quantum simulation. Beyond these architectures, various solid-state and photonic platforms are also being developed. While in the fault-tolerant era of computing, details of the underlying computing unit will not matter, in the near and intermediate terms, algorithm designers will take into account the pros and cons of hardware systems to increase efficiency.

%
\begin{figure}[t!]
\centering
\includegraphics[scale=0.625]{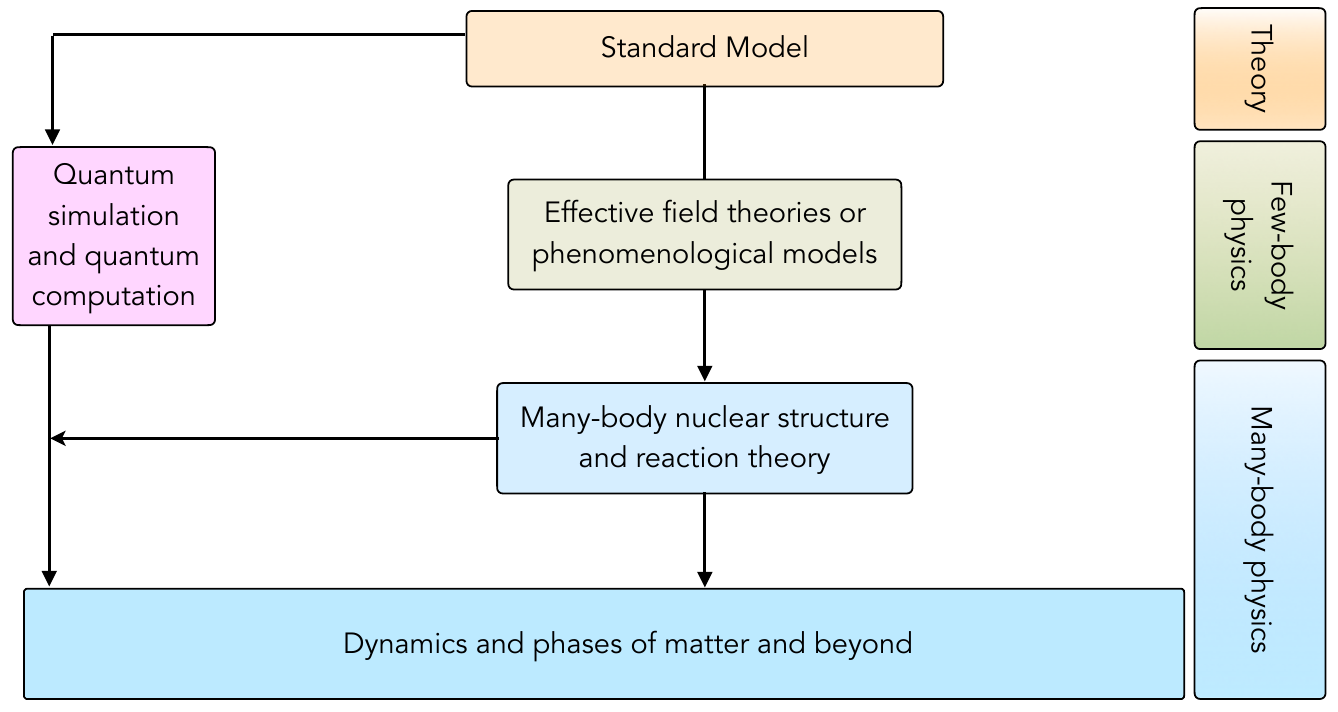}
\caption{A roadmap for leveraging quantum simulation and quantum computation in particle and nuclear physics.}
\label{fig:roadmap}
\end{figure}
Quantum computing can accelerate progress in particle and nuclear physics in at least two ways, as depicted in Fig.~\ref{fig:roadmap}. First, by encoding gauge-theory problems in quantum-computing language, one can directly attempt to access the dynamics and phases of matter from first-principles simulations. Such computations will be quantum and relativistic in nature, and are anticipated to be more demanding on resources. Second, by encoding effective field theories or phenomenological models of hadrons and nuclei in quantum computers, one can try to solve the nuclear many-body problem to access structure and reaction properties of nuclei and nuclear matter. Such computations are quantum and often nonrelativistic in nature, hence are expected to be less costly. While the latter case resembles quantum-chemistry and materials-science problems, the former case has nuances that need to be accounted for when developing algorithms and methods.

This talk aims to review the state of theory, algorithm, and hardware implementation and co-design for the gauge-theory simulation path. While early developments mostly focused on analog quantum simulation of simpler prototype models, it quickly became clear that analog simulation---in which the degrees of freedom and interactions in the hardware are mapped to the target physical system---will not be the best candidate for quantum simulation of the SM gauge theories, which exhibit a high degree of complexity. Universal digital quantum computing will likely be the only reliable path forward, although hybrid analog-digital approaches have gained momentum in recent years. This talk will, therefore, focus on digital and hybrid quantum computing for lattice gauge theories (LGTs). Basic elements of a quantum-computing-based lattice-field-theory program are introduced in  Sec.~\ref{sec:QC-LFT}; theoretical questions and pathways and solutions that address them are outlined in Sec.~\ref{sec:theory}; algorithmic progress in field-theory simulations is reviewed in  Sec.~\ref{sec:algorithm}, along with a discussion of the current cost estimates for quantum computing time dynamics in QCD; implementation, benchmark, and co-design frontiers of the program are reviewed in Sec.~\ref{sec:expt}, along with several experimental highlights; and finally a summary and outlook are provided in Sec.~\ref{sec:outlook}.

\section{A quantum-computing-based LFT program
\label{sec:QC-LFT}}

Quantum simulation generally involves three steps. First, nontrivial initial states need to be prepared on a quantum computer. For gauge-theory applications, such states include vacuum and hadronic states, thermal (equilibrium) states, and nonequilibrium states. Unfortunately, there is no proven quantum advantage in preparing a generic quantum state arbitrarily accurately, but heuristic algorithms, including hybrid classical-quantum variational ones, exist with reasonable performance. Often, a good trial state with nonvanishing overlap to the desired state can serve as the starting point for various algorithms with guaranteed success. Second, the states prepared may need to undergo nontrivial time dynamics. For example, hadronic wave packets evolve to collide and produce nontrivial final states, or nonequilibrium states may evolve into equilibrium states. Quantum algorithms for time evolution offer clear quantum advantage over known classical methods, presenting an opportunity for significant speedup. Third, properties of the resulting state need to be measured in an efficient manner, without demanding full state tomography (which is exponentially costly in system size). These properties include local and nonlocal, static and dynamical correlation functions, S-matrices and cross sections, or even measures of the state's entanglement structure, which can inform, e.g., thermalization and thermodynamic properties.
\begin{figure}[t!]
\centering
\includegraphics[scale=0.542]{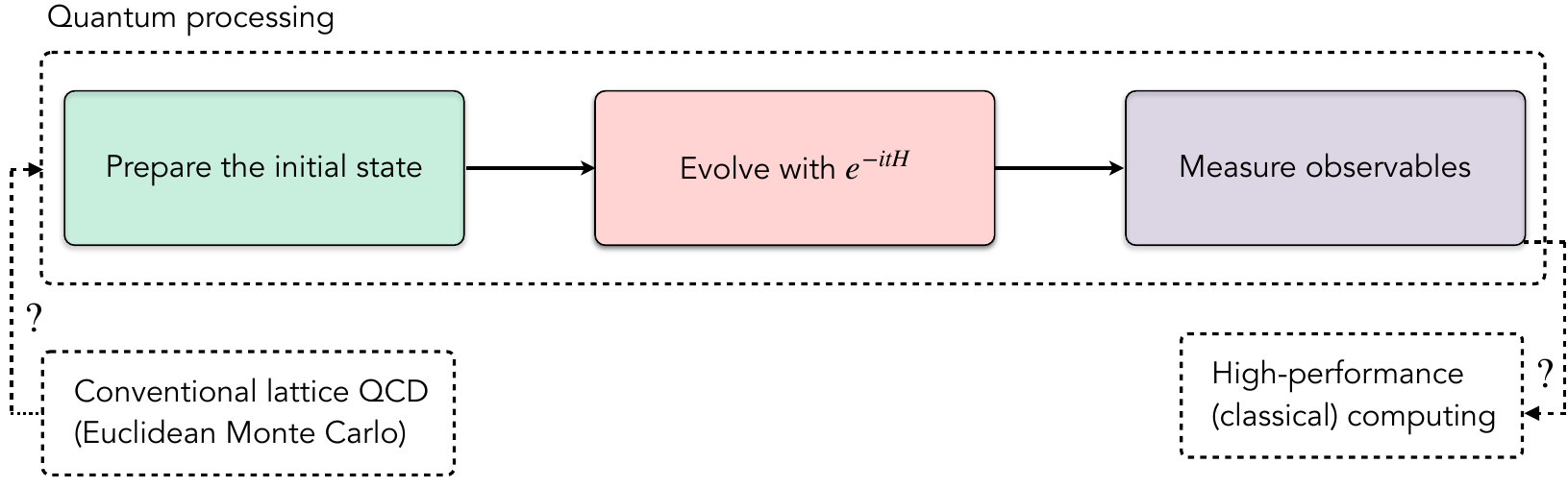}
\caption{Depicted are the quantum-simulation steps. Conventional lattice-QCD computations can inform and facilitate the state-preparation step, while high-performance (classical) computing will be needed to store and analyze vast quantum measurements. Time evolution can be accelerated by quantum processors, providing quantum advantage in many applications.}
\label{fig:roadmap}
\end{figure}
%

Classical HPC will likely still play a role in the quantum-computing era. For gauge-theory applications, one can envision a hybrid classical-quantum approach that involves accelerating state preparation on quantum computers using classical computations of various properties of the desired states and encoding them in an ansatz that has nonvanishing overlap to the state sought~\cite{Harmalkar:2020mpd,Gupta:2025xti}. Quantum processing units (QPUs) can continue implementing time evolution. Finally, the extensive amount of information encoded in the resulting states will likely need large, powerful classical HPC resources to be stored and processed. As a result, future HPC models will likely incorporate various processing units, from CPUs to GPUs to QPUs, with interconnects that translate classical and quantum input and output. Artificial intelligence and machine learning will likely play a role in such translations, providing classically efficient representations of states for easy access and recovery. Development of such models has already started in both industry and government sectors and will accelerate as fault-tolerant quantum computing moves closer to reality.

Various quantum algorithms exist for each step of the simulation. These algorithms inform the quantum operations required and the space and time scaling of resources as a function of system size, accuracy goal, and other simulation parameters. The algorithm design follows the simulation mode. There are at least three primary modes:
\begin{itemize}
\item[$\diamond$] Analog quantum simulation involves mapping the target system's degrees of freedom and Hamiltonian dynamics directly to those of the simulator, which often involves some degree of mode and Hamiltonian engineering. Once such a mapping is established, the system's evolution matches the simulator's evolution. When such a mapping exists, analog quantum simulation enables simulation of sizable system sizes and simulation times. While analog gauge-theory simulations have been successfully performed over the years, they remain limited to simpler models and lower dimensions (see Refs.~\cite{Aidelsburger:2021mia,Zohar:2021nyc,Halimeh:2023lid,Funcke:2023jbq,Halimeh:2025vvp} for recent reviews).
\item[$\diamond$] Digital quantum simulation often involves qubits---two-dimensional quantum modes---and amounts to performing time evolution and other tasks in discrete steps. These steps are further decomposed to a universal set of qubit operations---namely gates. For example, time evolution under local Hamiltonians can be Trotterized such that each Trotter step involves a product of evolution operators under each local Hamiltonian term separately. These operators are simpler to decompose into a gate set. Here, continuous time evolution is approximated up to an error that can be rigorously bounded and systematically reduced. Digital schemes are, therefore, flexible in simulating arbitrary Hamiltonian dynamics, including gauge theories, as demonstrated for models with varying complexity in recent years and reviewed in Sec.~\ref{sec:algorithm}. Moreover, quantum error correction and fault-tolerant computing conform to digital, gate-based computing.
\item[$\diamond$] Hybrid analog-digital schemes combine the best features of both analog and digital simulations for enhanced efficiency and flexibility. For example, when the target system's degrees of freedom can be mapped directly to intrinsic degrees of freedom in the simulator, such as fermions, bosons, or qudits, one may decompose digital evolution into intrinsic gate sets involving such degrees of freedom; and hence avoid qubit encoding to decrease the mapping overhead. Hybrid analog-digital implementations have recently been explored for field-theory simulations, as reviewed in Sec.~\ref{sec:expt}.
 \end{itemize}
 %

While many of the generic quantum algorithms for various simulation tasks are applicable to lattice field theories, several nuances remain. Theoretical efforts need to cast field theories in the Hamiltonian framework; algorithmic efforts need to translate the corresponding Hamiltonian dynamics to quantum circuits; and implementation, hardware benchmarks, and co-design need to follow to assess feasibility, performance, and empirical scalings. A quantum-simulation-based lattice-field-theory program will necessarily involve all these interconnected frontiers; indeed, lattice field theorists have been advancing the state-of-the-art on all these fronts. The remainder of this talk expands on the questions, requirements, and progress in these prongs.

\section{Theoretical frontiers
\label{sec:theory}}

What are the Hamiltonian formulations of the gauge theories of the Standard Model, and which formulations best suit the quantum-simulation tasks, particularly toward the continuum limit of LGTs? How does one truncate the infinite-dimensional Hilbert space of continuous gauge groups and how can the Hilbert-space truncation effects be estimated? What symmetries are broken in the truncated Hamiltonian formulations and how can one recover them? How can gauge invariance be best protected or restored? What are discretization and finite-volume effects on observables in the Minkowski-time Hamiltonian framework? How do these interplay with quantum-algorithmic errors such as time digitization and unitary synthesis errors? How can various observables, including scattering amplitudes, structure functions, or thermodynamical quantities, be reliably extracted from Minkowski-time Hamiltonians and correlation functions? What is the role of entanglement and measures of quantumness in the equilibrium and nonequilibrium physics of gauge theories, and can such information guide phenomenology? These theoretical questions and more continue to be studied and addressed.

The starting point of the quantum-simulation program is formulating a Hamiltonian framework. In the context of LGTs, the Hamiltonian consists of $H=H_I+H_M+H_E+H_B$. Here, $H_I$ denotes the fermion fields hopping between the lattice sites as they couple gauge invariantly to the gauge fields on lattice links; $H_M$ denotes the energy stored in the mass of fermions; $H_E$ and $H_B$ are associated with the energy stored in the electric and magnetic fields, respectively. A few remarks are in order:
\begin{itemize}
\item[$\diamond$] For continuous groups such as $U(1)$ and SU(N), the gauge-field Hilbert space at each link is infinite dimensional and needs to be truncated to allow finite-size computations. This truncation introduces a theoretical error that needs to be quantified and extrapolated away. Recent work attempts to improve the existing bounds on truncation errors by leveraging analytical tools~\cite{Tong:2021rfv}, Hilbert-space fragmentation~\cite{Ciavarella:2025tdl}, or Monte-Carlo improved energy-based methods~\cite{Yang:2026zpa}.
\item[$\diamond$] Only a fraction of the LGT Hilbert space is physical. Concretely, states that satisfy the Gauss's laws belong to the physical sector. While Hamiltonian dynamics is constrained to the physical Hilbert space, inexact algorithms and implementations can break gauge invariance and allow leakage into unphysical sectors. Alternatively, one may solve as many Gauss's laws as possible to reduce redundancy and Hilbert-space encoding cost, but the resulting Hamiltonian will be complex and nonlocal. How to balance redundant encoding and locality to optimize simulation cost is an important factor in designing algorithms.
\item[$\diamond$] The basis states chosen to represent the Hilbert space determine the structure of the Hamiltonian matrix and its computational complexity. In the electric and irreducible representation basis, $H_E$ is a diagonal operator while $H_I$ and $H_B$ are off-diagonal. In the conjugate basis or group-element basis, $H_E$ is off-diagonal while $H_I$ and $H_B$ are diagonal. The electric field is discrete but unbounded while the group elements are continuous but bounded. The former needs a cutoff on the electric-field value and the latter requires variable digitization, hence inducing different truncation errors.
\end{itemize}

The discussions above make it clear that various choices exist for formulating and representing the LGT Hamiltonian for simulation purposes. LGTs are not the only path to continuum field theories, and various finite-dimensional or qubitized models with the same continuum limit as the desired field theory are being examined. A vast landscape of options continues to be explored, constituting a vibrant research direction (see Refs.~\cite{Bauer:2023qgm,Bauer:2022hpo,Halimeh:2025vvp,Davoudi:2025kxb} for a few reviews).

\section{Algorithmic frontiers
\label{sec:algorithm}}
Can we develop near- and far-term algorithms with bounded errors and accurate resource requirements for gauge theories? Can given formulations/encodings of the Hamiltonian gauge theory reduce qubit and gate resources? Do approximate simulation algorithms break gauge invariance and how can they be protected efficiently? Can we leverage gauge-redundant encoding of LGTs to reduce error-correction overhead? Can interesting states be efficiently prepared on a quantum computer and how can nontrivial observables (such as scattering amplitudes, response and structure functions, or entanglement spectra) be measured? Progress on all these fronts has occurred, starting from the pioneering work of Byrnes and Yamamoto in the case of SU($N_c$) LGTs~\cite{Byrnes:2005qx}, and of Jordan, Lee, and Preskill~\cite{Jordan:2012xnu}, which provided a complete end-to-end algorithm for scattering of two particle wave packets in an interacting scalar field theory. Later, various extensions to gauge theories based on various encodings and formulations appeared (see e.g. Refs.~\cite{Byrnes:2005qx,Lamm:2019bik,Shaw:2020udc,Haase:2020kaj,Ciavarella:2021nmj,Kan:2021xfc,Davoudi:2022xmb,Murairi:2022zdg,Rhodes:2024zbr,Lamm:2024jnl,Balaji:2025afl,Halimeh:2025ivn,Ciavarella:2025tdl,Froland:2025bqf,davoudi2026qcd}). These works resulted in better understanding of the promise and reach of quantum algorithms in solving interesting problems in gauge theories. 

Given this progress, a natural question is: how much does it cost to simulate QCD dynamics? A complete resource estimate for digital quantum simulations of QCD remains an open problem, since the cost depends not only on the time-evolution task but also on state preparation, observable estimation, and control of finite-volume, lattice-spacing, truncation, digitization, measurement, and algorithmic errors. State preparation is particularly demanding for highly energetic hadrons, nuclei, scattering, and thermal states, and reliable estimates with bounded error and performance guarantees are not yet available. It is, therefore, useful to isolate real-time evolution, a central subroutine with rigorous algorithmic error bounds, and to focus on one set of parameters close to continuum and infinite-volume limits such that the results will be close to their physical limits.

For concreteness, let us consider the Kogut--Susskind Hamiltonian formulation in the electric-field basis, with a cutoff $\Lambda$ on the irrep quantum number~\cite{Kogut:1974ag,Byrnes:2005qx}. Here, implementing the exponentiated magnetic or plaquette operator, $H_B$, constitutes the dominant cost. The magnetic term contains the trace of products of four link operators around each plaquette.
\begin{itemize}
\item[$\diamond$] Any $2^n\times 2^n$-dimensional matrix operator can be decomposed to a sum of at most $4^n$ Pauli strings. A naive Pauli decomposition of the QCD Hamiltonian is, nonetheless, impractical. In the case of an $\mathrm{SU}(N_c)$ LGT, each link requires $O\left((N_c^2-1)\log\Lambda\right)$ qubits (with a binary encoding of gauge bosons into qubits), and a plaquette acts on four such registers, i.e., a $2^{4(N_c^2-1)\log\Lambda}\times2^{4(N_c^2-1)\log\Lambda}$-dimensional Hilbert space. The number of exponentiated Pauli strings to be implemented is then $O(\Lambda^{8(N_c^2-1)})$. For an $\mathrm{SU}(3)$ theory with a small cutoff $\Lambda=4$, the Pauli expansion contains roughly $4^{64}\sim10^{38}$ Pauli strings per plaquette. Moreover, Trotterization of these Pauli operators introduces significant Trotter error, hence requiring prohibitive cost to reach reasonable accuracies.

\item[$\diamond$] More efficient product-formula methods avoid full Pauli decomposition by implementing exponentials through block-diagonalization methods~\cite{Shaw:2020udc}. The number of terms can then be made independent of $\Lambda$, with only polylogarithmic cutoff dependence in matrix-element evaluation. A representative second-order product-formula gate-count scaling is~\cite{Kan:2021xfc}
\begin{align}
O\!\left(\nu^{3/2} V^{3/2} T^{3/2}\epsilon^{-1/2}\Lambda\,\mathrm{polylog}\!\left(V^{3/2}T^{3/2}\Lambda\epsilon^{-3/2}\right)
\right),
\nonumber
\end{align}
where $V$ is the spatial volume, $T$ the evolution time, $\epsilon$ the target accuracy on the norm of the time-evolution operator, and $\nu$ the number of implemented terms. Even with these improvements, concrete estimates for $V = (10\,\text{fm})^3$, $\epsilon$ ranging from $10^{-1}-10^{-3}$, lattice spacing $10^{-2}\,\text{fm}$, $\Lambda = 10$, and an evolution time comparable to the lattice extent, the required number of qubits and T gates are roughly $10^{11}$ and $10^{49}$--$10^{50}$, respectively. Further reduction of the above cost is possible using improved block-diagonalization methods. For example, using an algorithm based on singular-value-decomposition~\cite{davoudi2026qcd}, the number of terms to be implemented can be reduced by almost $9$ orders of magnitude, resulting in a reduction of almost $14$ orders of magnitude in the overall cost of the second-order product formula at a fixed accuracy. 

\item[$\diamond$] Post-Trotter methods provide a more favorable route in the fault-tolerant era of quantum computing. For example, a qubitization algorithm encodes the Hamiltonian directly into a block of a unitary matrix, so the plaquette contribution scales roughly as $O(N_c^4)$ terms (rather than exponentially in $N_c^2$ as in product-formula algorithms). Sparse-Hamiltonian techniques find and evaluate nonzero matrix elements on the fly, including Clebsch-Gordan coefficients, and a rotating frame with respect to the electric Hamiltonian avoids unfavorable cutoff scaling. With local fermion encodings, the gate complexity becomes~\cite{Rhodes:2024zbr}
\begin{align}
O\!\left(N_c^4VT\,\mathrm{polylog}(VT\epsilon^{-1})\right).
\nonumber
\end{align}
For the same representative setting outlined above, the time-evolution cost involves roughly the same qubit count but only $10^{27}$ T gates.
\end{itemize}
%
Digital QCD simulations, therefore, remain beyond near-term practicality. However, this discussion should not discourage progress for a few reasons. First, quantum simulation of smaller instances of QCD problems, and of larger instances of simpler Abelian and non-Abelian models, are fully within reach; many insights and predictions of phenomenological interest are, hence, anticipated. Second,  as the examples above demonstrate, the resource estimates will likely be further improved by continued algorithmic research. Third, the classical computational cost of QCD for static problems is comparable to the quantum costs above for dynamics problems, and the large qubit and gate counts should not come as a surprise. For example, recall that some of the largest QCD computations require fractions of exascale high-performance computing machines per year (amounting to roughly $10^{24}$ floating-point operations per calculation) and petabytes of storage (amounting to roughly $10^{16}$ classical bits). Future quantum supercomputers will hopefully arrive, and QCD problems will among the main use cases of the technology.

\section{Implementation, benchmark, and co-design frontiers
\label{sec:expt}}
\begin{figure}[t!]
\centering
\includegraphics[scale=0.461]{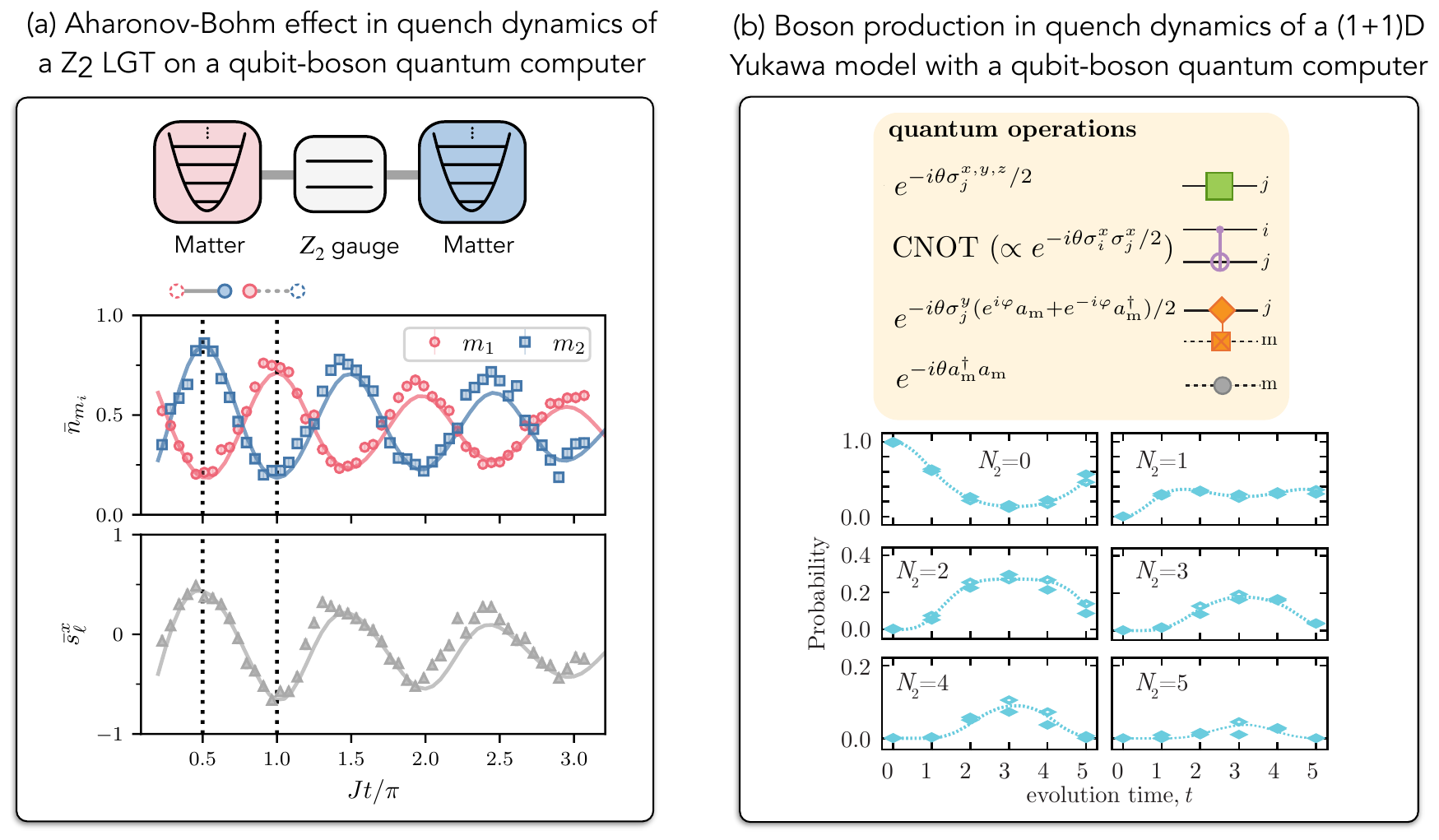}
\caption{Examples of the application of hybrid spin-boson quantum computation in select lattice field theories: (a) is adopted from Ref.~\cite{Saner:2025nrq} and (b) from Ref.~\cite{Than:2025gso}.}
\label{fig:hybrid}
\end{figure}

Lattice gauge theorists are some of the most active users of current quantum-hardware technology, and have performed some of the most complex quantum simulations to date. These studies follow a wealth of theoretical proposals for analog quantum simulation in the 2010s~\cite{Aidelsburger:2021mia,Zohar:2021nyc,Halimeh:2023lid,Halimeh:2025vvp} and were followed by a pioneering experiment on the Schwinger model pair-production dynamics in a four-qubit trapped-ion quantum computer~\cite{Martinez:2016yna}. This experiment, and many subsequent experiments, focused on creating nonequilibrium conditions after a quantum quench, in which an easily prepared non-eigenstate (such as a product state) undergoes nontrivial Hamiltonian dynamics. Phenomena such as string breaking, pair creation and hadronization, as well as hydrodynamization and thermalization can be studied in such experiments. Alternatively, energetic colliding hadronic states can be prepared to simulate the nonequilibrium aftermath of collision processes. Moreover, when such hadronic states are prepared, a wealth of dynamical correlation functions can be evaluated in these states, from which various nonperturbative structure and response functions can be extracted. Dense fermionic matter at given temperatures can also be prepared and its properties can be measured in equilibrium and nonequilibrium alike. Given the wealth and breadth of such studies, it will be impossible to review all results and implementations; here we focus on selected highlights, with an emphasis on results that have appeared in the past two years.
\begin{itemize}
\item[$\diamond$] \emph{Real-time dynamics}: An IBM quantum processor was used in Ref.~\cite{Mou:2025iiu} to simulate the quench dynamics of a (2+1)D U(1) LGT. Figure~\ref{fig:quench}(a) exhibits the electric field and charge expectation values after two Trotter steps of evolution in a one-plaquette system. Moreover, an IBM quantum processor was used in Ref.~\cite{Klco:2019evd} to probe the quench dynamics of a plaquette in a (2+1)D pure SU(2) LGT, as shown in Figure~\ref{fig:quench}(b). An IBM quantum processor was also used in Ref.~\cite{Ciavarella:2024fzw} in a (2+1)D pure SU(3) LGT at leading order in a Hamiltonian truncation method based on an expansion in $1/N_c$~\cite{Ciavarella:2024fzw,Ciavarella:2025bsg}. Figure~\ref{fig:quench}(b) exhibits the Trotterized evolution of a plaquette expectation value as a function of Minkowski time in multi-plaquette systems.
\item[$\diamond$] \emph{Thermalization dynamics}: A trapped-ion quantum computer at Duke University was used in Ref.~\cite{Mueller:2024mmk} to probe thermalization stages of a (2+1)D $Z_2$ LGT in the quench dynamics of a 10-plaquette system using entanglement spectroscopy of a 4-plaquette subsystem. Specifically, as shown in Fig.~\ref{fig:thermalization}, statistical properties of the entanglement spectrum, including gap-ratio distribution, at early, intermediate, and late stages of evolution reveal an approach to Gaussian-unitary-ensemble statistics, exhibiting chaos, an early sign of thermalization.
\begin{figure}[b!]
\centering
\includegraphics[scale=0.4225]{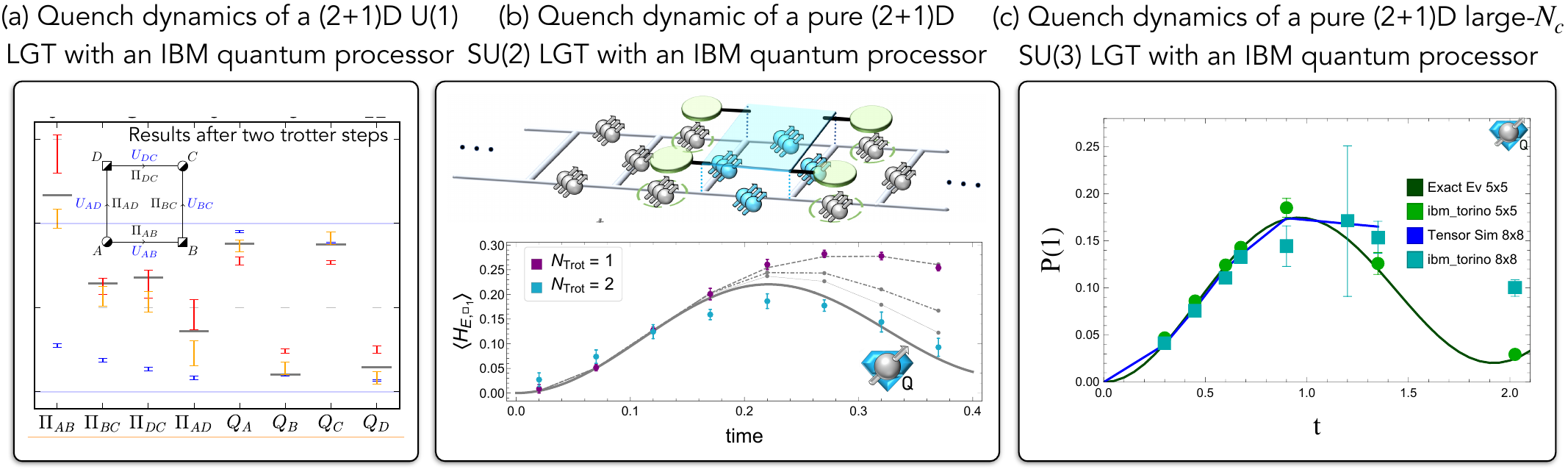}
\caption{Examples of nonequilibrium quench dynamics in select (2+1)D LGTs: (a) is adopted from Ref.~\cite{Mou:2025iiu}, (b) from Ref.~\cite{Klco:2019evd}, and (c) from Ref.~\cite{Ciavarella:2024fzw}.}
\label{fig:quench}
\end{figure}
\begin{figure}[t!]
\centering
\includegraphics[scale=0.6]{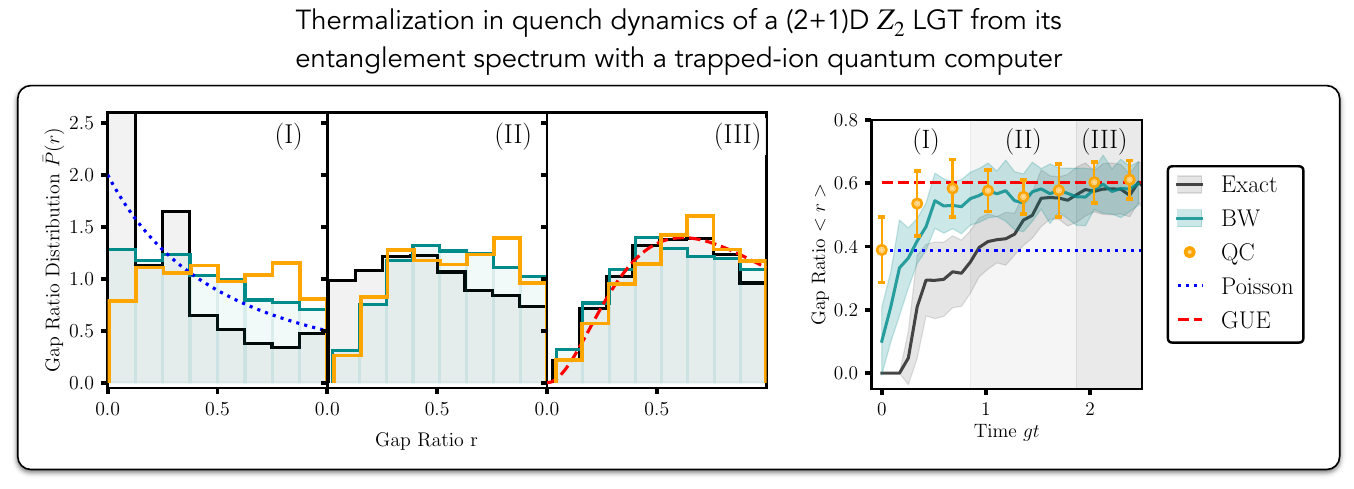}
\caption{Stages of thermalization dynamics in a (2+1)D $Z_2$ LGT learned from the nonequilibrium state's entanglement spectrum. Plots are adopted from Ref.~\cite{Mueller:2024mmk}.}
\label{fig:thermalization}
\end{figure}
\begin{figure}[t!]
\centering
\includegraphics[scale=0.61]{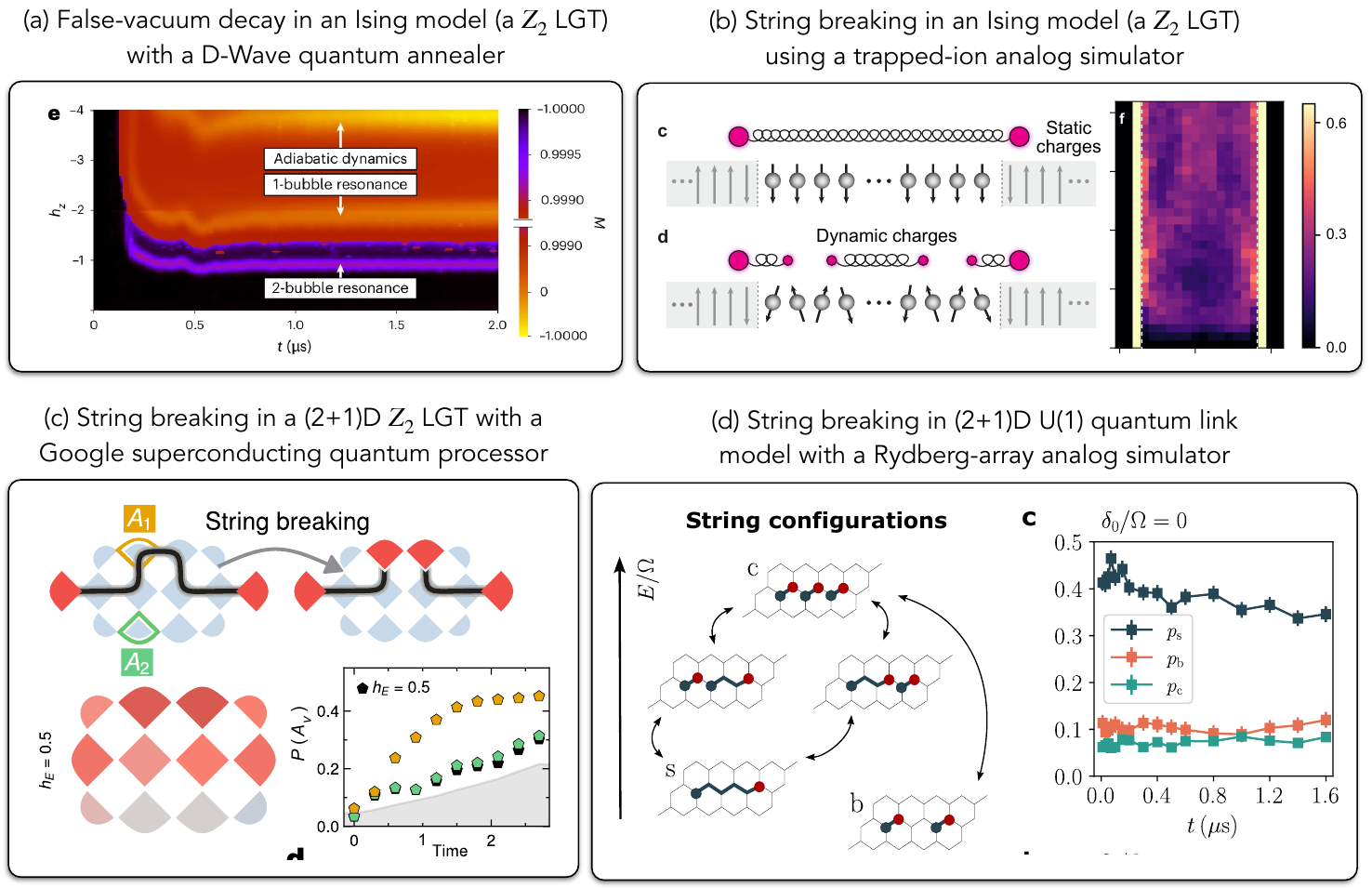}
\caption{Examples of false-vacuum decay and string breaking in quench dynamics in select LGTs: (a) is adopted from Ref.~\cite{Vodeb:2024tvo}, (b) from Ref.~\cite{De:2024smi}, (c) from Ref.~\cite{Cochran:2024rwe}, and (d) from Ref.~\cite{Gonzalez-Cuadra:2024xul}.}
\label{fig:string}
\end{figure}
\item[$\diamond$] \emph{String breaking}: Ising spin chains can have a dual LGT and have been used as a playground for simulations of confinement and vacuum dynamics. For example, a D-Wave quantum annealer was used in Ref.~\cite{Vodeb:2024tvo} to probe bubble-nucleation dynamics in the decay of a false vacuum near $n$-bubble resonances, as shown in Fig.~\ref{fig:string}(a). Additionally, the quench dynamics of a long string excitation was probed in Ref.~\cite{De:2024smi} using an analog trapped-ion quantum simulator at Duke University, revealing particle production and propagation near the string edges, as shown in Fig.~\ref{fig:string}(b). String breaking in a (2+1)D $Z_2$ LGT was also probed in Ref.~\cite{Cochran:2024rwe} in the quench dynamics of a six-plaquette system using a Google quantum processor, with representative plots shown in Fig.~\ref{fig:string}(c). Such dynamics in a (1+1)D U(1) quantum link model were also probed in Ref.~\cite{Gonzalez-Cuadra:2024xul} via mapping to a 2D Rydberg-array analog quantum simulator by QuEra, with representative plots shown in Fig.~\ref{fig:string}(d).
\item[$\diamond$] \emph{Collider observables}: Among nonperturbative quantities of relevance to collider physics accessible to quantum computers is a parton distribution function. This quantity was computed in the lattice Schwinger model in Ref.~\cite{Chen:2025zeh} using an IBM quantum processor, as shown in Fig.~\ref{fig:pdf}(a). Another set of nonperturbative quantities are energy-energy correlators, of relevance to elucidating jet substructure in colliders. One such correlator was computed in a (2+1)D SU(2) LGT in Ref.~\cite{Lee:2024jnt} using an IBM quantum processor, as shown in Fig.~\ref{fig:pdf}(b).
\begin{figure}[t!]
\centering
\includegraphics[scale=0.6]{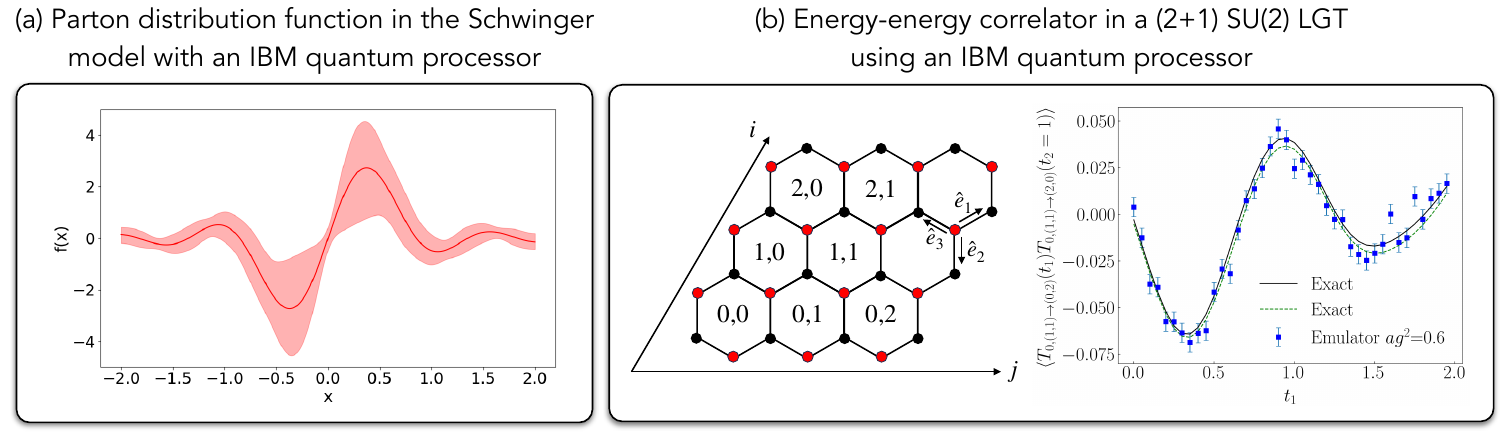}
\caption{Examples of a parton distribution function and an energy-energy correlator in select LGTs: (a) is adopted from Ref.~\cite{Chen:2025zeh} and (b) from Ref.~\cite{Lee:2024jnt}.}
\label{fig:pdf}
\end{figure}
\item[$\diamond$] \emph{Scattering and transitions}: Following progress in preparation of hadronic wave packets on quantum hardware~\cite{Farrell:2024fit,Davoudi:2024wyv}, computations of two-particle scattering dynamics have emerged. For example, fermion-antifermion scattering in a (1+1)D Thirring model was simulated in Ref.~\cite{Chai:2023qpq} with an IBM quantum processor [Fig.~\ref{fig:scattering}(a)]; scattering in a (1+1)D Ising field theory was simulated in Ref.~\cite{Farrell:2025nkx} with an IBM quantum processor, with inelastic components identified in the long-time dynamics using a circuit emulator [Fig.~\ref{fig:scattering}(b)]; two-hadron scattering in a (1+1)D U(1) quantum link model was achieved in Ref.~\cite{Schuhmacher:2025ehh} with an IBM quantum processor, where long-time simulation made possible by simplified initial states [Fig.~\ref{fig:scattering}(c)]; and two-hadron scattering in a (1+1)D $Z_2$ LGT was achieved in Ref.~\cite{Davoudi:2025rdv} with an IonQ quantum computer, with an accurate but resource-demanding wave-packet preparation step [Fig.~\ref{fig:scattering}(d)]. Moreover, a demonstration of real-time transition dynamics of (1+1)D QCD coupled to leptons was presented in Ref.~\cite{Farrell:2022vyh} for a $\beta$ decay transition with a Quantinuum quantum computer [Fig.~\ref{fig:beta}(a)], and in Ref.~\cite{Chernyshev:2025lil} for a neutrinoless double-$\beta$ decay transition with an IonQ quantum computer [Fig.~\ref{fig:beta}(b)].
\begin{figure}[t!]
\centering
\includegraphics[scale=0.5525]{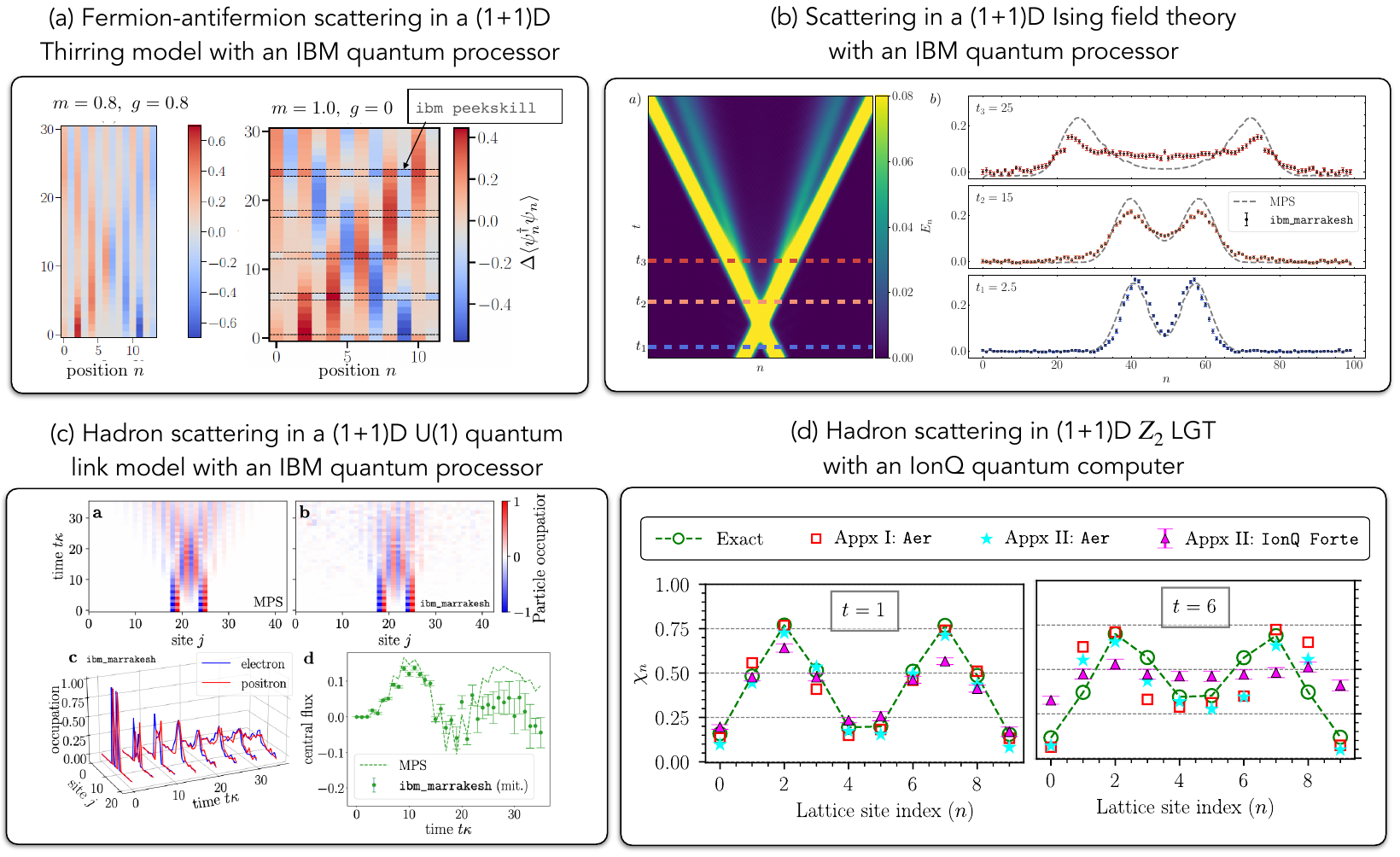}
\caption{Examples of two-particle scattering in select LGTs: (a) is adopted from Ref.~\cite{Chai:2023qpq}, (b) from Ref.~\cite{Farrell:2025nkx}, (c) from Ref.~\cite{Schuhmacher:2025ehh}, and (d) from Ref.~\cite{Davoudi:2025rdv}.}
\label{fig:scattering}
\end{figure}
\begin{figure}[t!]
\centering
\includegraphics[scale=0.552]{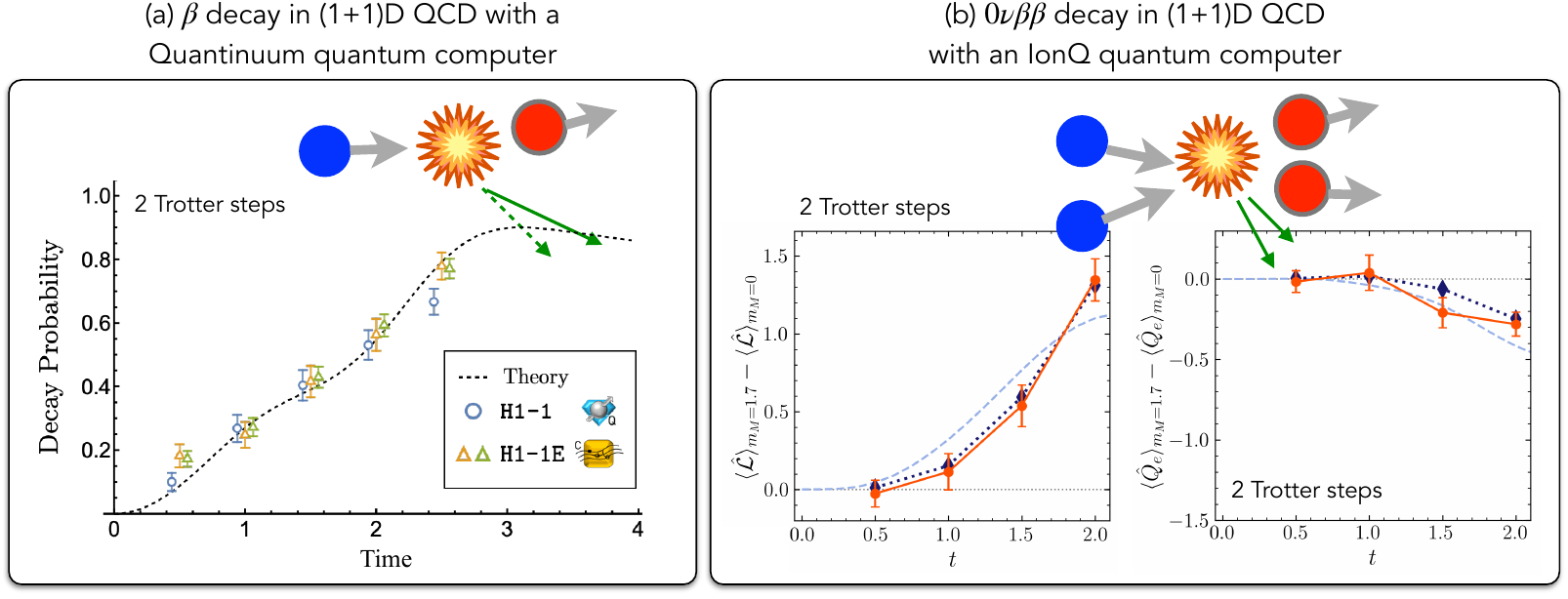}
\caption{Single- and double-$\beta$ decay probability amplitudes in (1+1)D QCD: (a) is adopted from Ref.~\cite{Farrell:2022vyh} and (b) from Ref.~\cite{Chernyshev:2025lil}.}
\label{fig:beta}
\end{figure}
\item[$\diamond$] \emph{Phase diagrams}: The phase diagram of a (2+1) U(1) LGT coupled to two flavors of fermions was computed in Ref.~\cite{Rosanowski:2025nck} using an IBM quantum processor. As shown in Fig.~\ref{fig:phase}(a), the fermion-number difference as a function of the chemical potential signifies a discrete change. Another study~\cite{Than:2024zaj} used a trapped-ion quantum computer at the University of Maryland to prepare the thermal state of fermions in a two-site (1+1)D QCD at finite chemical potential. As shown in Fig.~\ref{fig:phase}(b), the chiral condensate as a function of the chemical potential signifies a rapid change.
\begin{figure}[t!]
\centering
\includegraphics[scale=0.456]{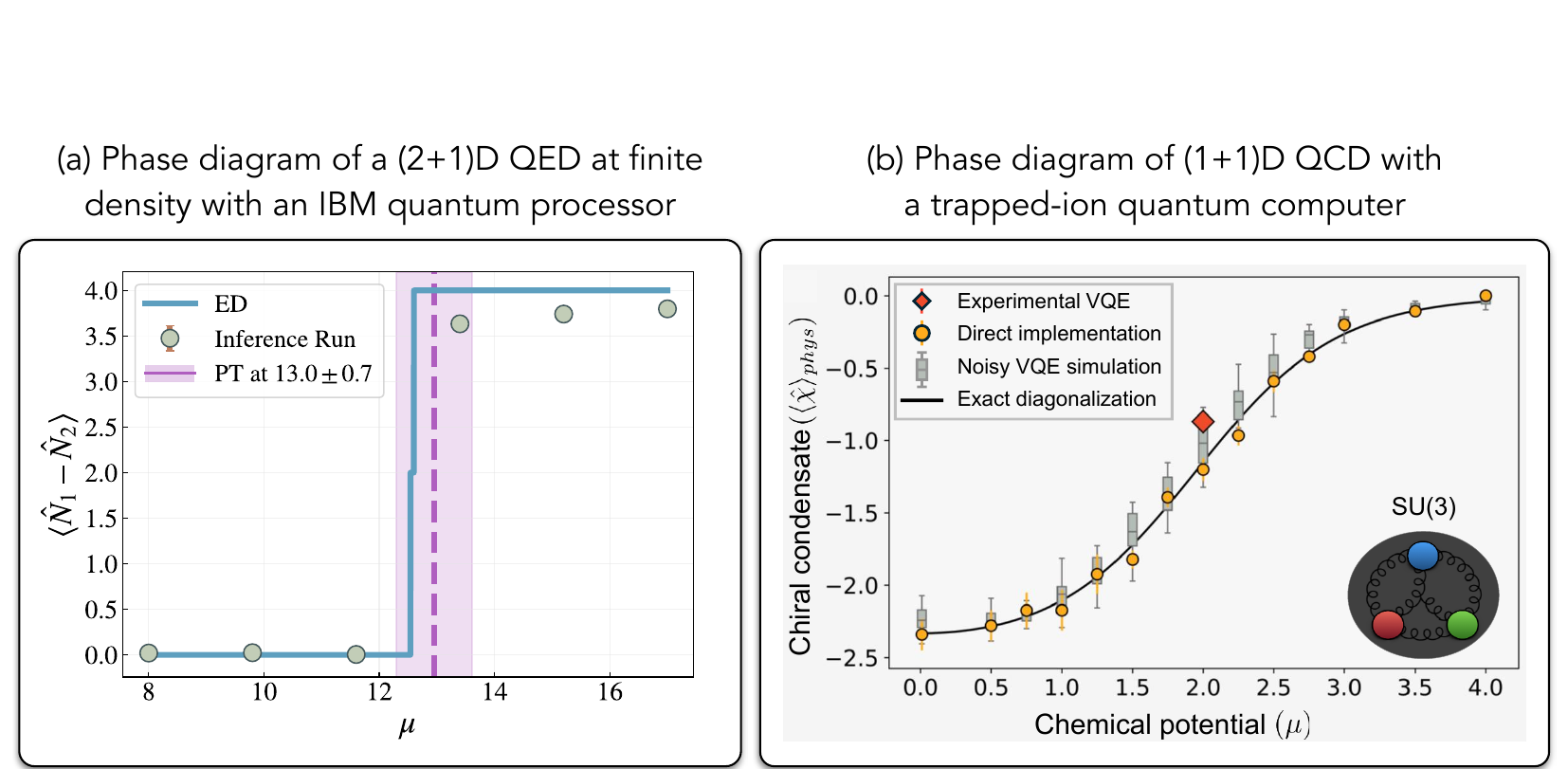}
\caption{Examples of phase diagrams at finite density in select LGTs: (a) is adopted from Ref.~\cite{Rosanowski:2025nck} and (b) from Ref.~\cite{Than:2024zaj}.}
\label{fig:phase}
\end{figure}
\end{itemize}
%
Finally, it is conceivable that the quantum-computing developments will be largely influenced by the applications' specifics and requirements. In fact, such co-design efforts have already taken place and have led to advances in hardware design and simulation implementations. For example, trapped ions have shown unique capabilities as a hybrid analog-digital computer. In fact, lattice-field-theory simulations have driven the use of phonon modes as dynamical degrees of freedom on quantum hardware. As shown in Fig.~\ref{fig:hybrid}(a), quench dynamics of a (1+1)D $Z_2$ LGT was enabled in Ref.~\cite{Saner:2025nrq} by mapping the fermion fields onto the two lowest phonon states and the gauge bosons onto the qubits. Furthermore, as shown in Fig.~\ref{fig:hybrid}(b), using a phonon-qubit gate set~\cite{Davoudi:2021ney}, the Trotterized evolution of a (1+1)D Yukawa field theory was simulated in Ref.~\cite{Than:2025gso}, in which the scalar fields were encoded in phonon excitations and the fermion fields in the qubits. Such boson-to-boson mappings will greatly reduce the computational cost of bosonic field theories and represent an exciting frontier for further developments. Related efforts for gauge-theory simulations have also been proposed for other hardware platforms, including superconducting circuits and cavity QED~\cite{Crane:2024tlj,Popov:2023xft,Illa:2024kmf,Kurkcuoglu:2024cfv}, Rydberg atoms~\cite{Gonzalez-Cuadra:2022hxt}, and qudit-based trapped ions~\cite{Meth:2023wzd} (with a first implementation reported in the latter).

\section{Summary and outlook
\label{sec:outlook}}

Quantum simulation is emerging as a natural extension of the lattice-field-theory program in particle and nuclear physics. Conventional Euclidean Monte Carlo methods have delivered major successes for static hadronic and nuclear observables, but they remain limited for large atomic nuclei, dense matter, and general real-time observables due to signal-to-noise and sign problems. Quantum computers offer a complementary route: states can be prepared, evolved directly under Hamiltonian dynamics, and measured to obtain observables tailored to various physics targets. A plausible long-term path for quantum simulation of the Standard Model gauge theories is digital, fault-tolerant quantum computation; nonetheless, hybrid analog-digital approaches may reduce overhead by exploiting native bosonic, fermionic, or qudit degrees of freedom in quantum hardware.

Progress has been made across theory, algorithms, and hardware. Theoretical work has led to the development of Hamiltonian formulations with finite-dimensional truncations of continuous gauge fields; incorporation and protection of gauge invariance; analyses of continuum and finite-volume systematics; and schemes for obtaining scattering amplitudes, structure and response functions, and transport, thermodynamic, and entanglement quantities. Algorithmic advances have moved beyond naive Pauli decompositions and Trotterization, while physics-inspired circuit constructions promise to reduce projected costs. Experiments on superconducting, trapped-ion, and Rydberg-array platforms have so far demonstrated real-time quench dynamics, string breaking, thermalization signatures, scattering processes, transition dynamics, and phase-diagram calculations in lower-dimensional or truncated LGTs.

Full QCD simulations with controlled systematics remain beyond current hardware, but near-term progress will shed light on many unexplored phenomenological aspects of finite-density and dynamical processes in gauge theories. The upcoming studies will focus on controlled benchmark problems, more complex Abelian and non-Abelian models, larger system sizes in higher dimensions, tighter resource estimates, and co-design between algorithms and hardware. Classical high-performance computing will remain essential for state construction, measurement-data storage, and physics analysis, constituting a hybrid classical-quantum enterprise. Such developments make quantum simulation a powerful, complementary route to nonperturbative studies in particle and nuclear physics.

\section*{Acknowledgements}
Z.D. acknowledges support by the U.S. Department of Energy (DOE), Office of Science, Office of Nuclear Physics (award no. DE-SC0026067);  the U.S. National Science Foundation’s Quantum Leap Challenge Institute (award no. OMA-2120757); the DOE, Office of Science, Office of Advanced Scientific Computing Research (ASCR), program in Accelerated Research in Quantum Computing, Fundamental Algorithmic Research toward Quantum Utility (FARQu); and the Maryland Center for Fundamental Physics, Department of Physics, and College of Computer, Mathematical, and Natural Sciences at the University of Maryland. Z.D. is further grateful to the Max Planck Institute for Quantum Optics (MPQ), Garching, Germany, and the Institute for Nuclear Theory (INT), Seattle, USA, for their hospitality during the completion of these proceedings. The visit to MPQ was made possible by a Humboldt Research Fellowship and to the INT by the U.S. DOE (award no. No. DE-FG02-00ER41132).

\bibliographystyle{JHEP}
\bibliography{bibi}

\providecommand{\href}[2]{#2}\begingroup\raggedright\begin{thebibliography}{10}

\bibitem{FlavourLatticeAveragingGroupFLAG:2024oxs}
{\scshape Flavour Lattice Averaging Group (FLAG)} collaboration,
  {}\href{https://doi.org/10.1103/nfzp-p5dn}{\emph{Phys. Rev. D} {\bfseries
  113} (2026) 014508} [\href{https://arxiv.org/abs/2411.04268}{{\ttfamily
  2411.04268}}].

\bibitem{Davoudi:2022bnl}
Z.~Davoudi et~al., {} in \emph{{Snowmass 2021}}, 9, 2022
  [\href{https://arxiv.org/abs/2209.10758}{{\ttfamily 2209.10758}}].

\bibitem{Luscher:1986pf}
M.~Luscher, {}\href{https://doi.org/10.1007/BF01211097}{\emph{Commun. Math.
  Phys.} {\bfseries 105} (1986) 153}.

\bibitem{Ji:2013dva}
X.~Ji, {}\href{https://doi.org/10.1103/PhysRevLett.110.262002}{\emph{Phys. Rev.
  Lett.} {\bfseries 110} (2013) 262002}
  [\href{https://arxiv.org/abs/1305.1539}{{\ttfamily 1305.1539}}].

\bibitem{Ryu:2006bv}
S.~Ryu and T.~Takayanagi,
  {}\href{https://doi.org/10.1103/PhysRevLett.96.181602}{\emph{Phys. Rev.
  Lett.} {\bfseries 96} (2006) 181602}
  [\href{https://arxiv.org/abs/hep-th/0603001}{{\ttfamily hep-th/0603001}}].

\bibitem{VanRaamsdonk:2010pw}
M.~Van~Raamsdonk, {}\href{https://doi.org/10.1007/s10714-010-1034-0}{\emph{Gen.
  Rel. Grav.} {\bfseries 42} (2010) 2323}
  [\href{https://arxiv.org/abs/1005.3035}{{\ttfamily 1005.3035}}].

\bibitem{Kitaev:2005dm}
A.~Kitaev and J.~Preskill,
  {}\href{https://doi.org/10.1103/PhysRevLett.96.110404}{\emph{Phys. Rev.
  Lett.} {\bfseries 96} (2006) 110404}
  [\href{https://arxiv.org/abs/hep-th/0510092}{{\ttfamily hep-th/0510092}}].

\bibitem{Levin:2006zz}
M.~Levin and X.-G.~Wen,
  {}\href{https://doi.org/10.1103/PhysRevLett.96.110405}{\emph{Phys. Rev.
  Lett.} {\bfseries 96} (2006) 110405}.

\bibitem{Li:2008kda}
H.~Li and F.D.M.~Haldane,
  {}\href{https://doi.org/10.1103/PhysRevLett.101.010504}{\emph{Phys. Rev.
  Lett.} {\bfseries 101} (2008) 010504}
  [\href{https://arxiv.org/abs/0805.0332}{{\ttfamily 0805.0332}}].

\bibitem{Mueller:2021gxd}
N.~Mueller, T.V.~Zache and R.~Ott,
  {}\href{https://doi.org/10.1103/PhysRevLett.129.011601}{\emph{Phys. Rev.
  Lett.} {\bfseries 129} (2022) 011601}
  [\href{https://arxiv.org/abs/2107.11416}{{\ttfamily 2107.11416}}].

\bibitem{Davoudi:2024osg}
Z.~Davoudi, C.~Jarzynski, N.~Mueller, G.~Oruganti, C.~Powers and N.Y.~Halpern,
  {}\href{https://doi.org/10.1103/PhysRevLett.133.250402}{\emph{Phys. Rev.
  Lett.} {\bfseries 133} (2024) 250402}
  [\href{https://arxiv.org/abs/2404.02965}{{\ttfamily 2404.02965}}].

\bibitem{Feynman:1982}
R.P.~Feynman, {}\href{https://doi.org/10.1007/BF02650179}{\emph{International
  Journal of Theoretical Physics} {\bfseries 21} (1982) 467}.

\bibitem{Lloyd:1996}
S.~Lloyd, {}\href{https://doi.org/10.1126/science.273.5278.1073}{\emph{Science}
  {\bfseries 273} (1996) 1073}.

\bibitem{Altman:2019vbv}
E.~Altman et~al.,
  {}\href{https://doi.org/10.1103/PRXQuantum.2.017003}{\emph{PRX Quantum}
  {\bfseries 2} (2021) 017003}
  [\href{https://arxiv.org/abs/1912.06938}{{\ttfamily 1912.06938}}].

\bibitem{Preskill:2018jim}
J.~Preskill, {}\href{https://doi.org/10.22331/q-2018-08-06-79}{\emph{Quantum}
  {\bfseries 2} (2018) 79} [\href{https://arxiv.org/abs/1801.00862}{{\ttfamily
  1801.00862}}].

\bibitem{Regent:2025cos}
F.-M.L.~R{\'e}gent, {} \href{https://arxiv.org/abs/2507.03678}{{\ttfamily
  2507.03678}}.

\bibitem{Harmalkar:2020mpd}
S.~Harmalkar, H.~Lamm and S.~Lawrence.

\bibitem{Gupta:2025xti}
N.~Gupta, C.D.~White and Z.~Davoudi, {}
  \href{https://arxiv.org/abs/2506.02313}{{\ttfamily 2506.02313}}.

\bibitem{Aidelsburger:2021mia}
M.~Aidelsburger et~al.,
  {}\href{https://doi.org/10.1098/rsta.2021.0064}{\emph{Phil. Trans. Roy. Soc.
  Lond. A} {\bfseries 380} (2021) 20210064}
  [\href{https://arxiv.org/abs/2106.03063}{{\ttfamily 2106.03063}}].

\bibitem{Zohar:2021nyc}
E.~Zohar, {}\href{https://doi.org/10.1098/rsta.2021.0069}{\emph{Phil. Trans.
  Roy. Soc. Lond. A} {\bfseries 380} (2022) 20210069}
  [\href{https://arxiv.org/abs/2106.04609}{{\ttfamily 2106.04609}}].

\bibitem{Halimeh:2023lid}
J.C.~Halimeh, M.~Aidelsburger, F.~Grusdt, P.~Hauke and B.~Yang,
  {}\href{https://doi.org/10.1038/s41567-024-02721-8}{\emph{Nature Phys.}
  {\bfseries 21} (2025) 25} [\href{https://arxiv.org/abs/2310.12201}{{\ttfamily
  2310.12201}}].

\bibitem{Funcke:2023jbq}
L.~Funcke, T.~Hartung, K.~Jansen and S.~K{\"u}hn,
  {}\href{https://doi.org/10.22323/1.430.0228}{\emph{PoS} {\bfseries
  LATTICE2022} (2023) 228} [\href{https://arxiv.org/abs/2302.00467}{{\ttfamily
  2302.00467}}].

\bibitem{Halimeh:2025vvp}
J.C.~Halimeh, N.~Mueller, J.~Knolle, Z.~Papi{\'c} and Z.~Davoudi, {}
  \href{https://arxiv.org/abs/2509.03586}{{\ttfamily 2509.03586}}.

\bibitem{Tong:2021rfv}
Y.~Tong, V.V.~Albert, J.R.~McClean, J.~Preskill and Y.~Su,
  {}\href{https://doi.org/10.22331/q-2022-09-22-816}{\emph{Quantum} {\bfseries
  6} (2022) 816} [\href{https://arxiv.org/abs/2110.06942}{{\ttfamily
  2110.06942}}].

\bibitem{Ciavarella:2025tdl}
A.N.~Ciavarella, S.~Hariprakash, J.C.~Halimeh and C.W.~Bauer, {}
  \href{https://arxiv.org/abs/2508.00061}{{\ttfamily 2508.00061}}.

\bibitem{Yang:2026zpa}
J.~Yang, C.F.~Kane and S.~Jabeen, {}
  \href{https://arxiv.org/abs/2604.24896}{{\ttfamily 2604.24896}}.

\bibitem{Bauer:2023qgm}
C.W.~Bauer, Z.~Davoudi, N.~Klco and M.J.~Savage,
  {}\href{https://doi.org/10.1038/s42254-023-00599-8}{\emph{Nature Rev. Phys.}
  {\bfseries 5} (2023) 420} [\href{https://arxiv.org/abs/2404.06298}{{\ttfamily
  2404.06298}}].

\bibitem{Bauer:2022hpo}
C.W.~Bauer et~al.,
  {}\href{https://doi.org/10.1103/PRXQuantum.4.027001}{\emph{PRX Quantum}
  {\bfseries 4} (2023) 027001}
  [\href{https://arxiv.org/abs/2204.03381}{{\ttfamily 2204.03381}}].

\bibitem{Davoudi:2025kxb}
Z.~Davoudi, {} \href{https://arxiv.org/abs/2507.15840}{{\ttfamily 2507.15840}}.

\bibitem{Byrnes:2005qx}
T.~Byrnes and Y.~Yamamoto,
  {}\href{https://doi.org/10.1103/PhysRevA.73.022328}{\emph{Phys. Rev. A}
  {\bfseries 73} (2006) 022328}
  [\href{https://arxiv.org/abs/quant-ph/0510027}{{\ttfamily
  quant-ph/0510027}}].

\bibitem{Jordan:2012xnu}
S.P.~Jordan, K.S.M.~Lee and J.~Preskill,
  {}\href{https://doi.org/10.1126/science.1217069}{\emph{Science} {\bfseries
  336} (2012) 1130} [\href{https://arxiv.org/abs/1111.3633}{{\ttfamily
  1111.3633}}].

\bibitem{Lamm:2019bik}
H.~Lamm, S.~Lawrence and Y.~Yamauchi,
  {}\href{https://doi.org/10.1103/PhysRevD.100.034518}{\emph{Phys. Rev. D}
  {\bfseries 100} (2019) 034518}
  [\href{https://arxiv.org/abs/1903.08807}{{\ttfamily 1903.08807}}].

\bibitem{Shaw:2020udc}
A.F.~Shaw, P.~Lougovski, J.R.~Stryker and N.~Wiebe,
  {}\href{https://doi.org/10.22331/q-2020-08-10-306}{\emph{Quantum} {\bfseries
  4} (2020) 306} [\href{https://arxiv.org/abs/2002.11146}{{\ttfamily
  2002.11146}}].

\bibitem{Haase:2020kaj}
J.F.~Haase, L.~Dellantonio, A.~Celi, D.~Paulson, A.~Kan, K.~Jansen et~al.,
  {}\href{https://doi.org/10.22331/q-2021-02-04-393}{\emph{Quantum} {\bfseries
  5} (2021) 393} [\href{https://arxiv.org/abs/2006.14160}{{\ttfamily
  2006.14160}}].

\bibitem{Ciavarella:2021nmj}
A.~Ciavarella, N.~Klco and M.J.~Savage,
  {}\href{https://doi.org/10.1103/PhysRevD.103.094501}{\emph{Phys. Rev. D}
  {\bfseries 103} (2021) 094501}
  [\href{https://arxiv.org/abs/2101.10227}{{\ttfamily 2101.10227}}].

\bibitem{Kan:2021xfc}
A.~Kan and Y.~Nam, {} \href{https://arxiv.org/abs/2107.12769}{{\ttfamily
  2107.12769}}.

\bibitem{Davoudi:2022xmb}
Z.~Davoudi, A.F.~Shaw and J.R.~Stryker,
  {}\href{https://doi.org/10.22331/q-2023-12-20-1213}{\emph{Quantum} {\bfseries
  7} (2023) 1213} [\href{https://arxiv.org/abs/2212.14030}{{\ttfamily
  2212.14030}}].

\bibitem{Murairi:2022zdg}
E.M.~Murairi, M.J.~Cervia, H.~Kumar, P.F.~Bedaque and A.~Alexandru,
  {}\href{https://doi.org/10.1103/PhysRevD.106.094504}{\emph{Phys. Rev. D}
  {\bfseries 106} (2022) 094504}
  [\href{https://arxiv.org/abs/2208.11789}{{\ttfamily 2208.11789}}].

\bibitem{Rhodes:2024zbr}
M.L.~Rhodes, M.~Kreshchuk and S.~Pathak,
  {}\href{https://doi.org/10.1103/PRXQuantum.5.040347}{\emph{PRX Quantum}
  {\bfseries 5} (2024) 040347}
  [\href{https://arxiv.org/abs/2405.10416}{{\ttfamily 2405.10416}}].

\bibitem{Lamm:2024jnl}
H.~Lamm, Y.-Y.~Li, J.~Shu, Y.-L.~Wang and B.~Xu,
  {}\href{https://doi.org/10.1103/PhysRevD.110.054505}{\emph{Phys. Rev. D}
  {\bfseries 110} (2024) 054505}
  [\href{https://arxiv.org/abs/2405.12890}{{\ttfamily 2405.12890}}].

\bibitem{Balaji:2025afl}
P.~Balaji, C.~Conefrey-Shinozaki, P.~Draper, J.K.~Elhaderi, D.~Gupta,
  L.~Hidalgo et~al., {}\href{https://doi.org/10.1103/k8f6-yft8}{\emph{Phys.
  Rev. D} {\bfseries 112} (2025) 054511}
  [\href{https://arxiv.org/abs/2503.08866}{{\ttfamily 2503.08866}}].

\bibitem{Halimeh:2025ivn}
J.C.~Halimeh, M.~Hanada and S.~Matsuura, {}
  \href{https://arxiv.org/abs/2506.18966}{{\ttfamily 2506.18966}}.

\bibitem{Froland:2025bqf}
H.~Froland, D.M.~Grabowska and Z.~Li, {}
  \href{https://arxiv.org/abs/2512.22782}{{\ttfamily 2512.22782}}.

\bibitem{davoudi2026qcd}
Z.~Davoudi and J.~Stryker, {}{\emph{in preparation} (2026) }.

\bibitem{Kogut:1974ag}
J.B.~Kogut and L.~Susskind,
  {}\href{https://doi.org/10.1103/PhysRevD.11.395}{\emph{Phys. Rev. D}
  {\bfseries 11} (1975) 395}.

\bibitem{Martinez:2016yna}
E.A.~Martinez et~al.,
  {}\href{https://doi.org/10.1038/nature18318}{\emph{Nature} {\bfseries 534}
  (2016) 516} [\href{https://arxiv.org/abs/1605.04570}{{\ttfamily
  1605.04570}}].

\bibitem{Mou:2025iiu}
Z.-G.~Mou and B.~Chakraborty, {}
  \href{https://arxiv.org/abs/2510.27668}{{\ttfamily 2510.27668}}.

\bibitem{Klco:2019evd}
N.~Klco, J.R.~Stryker and M.J.~Savage,
  {}\href{https://doi.org/10.1103/PhysRevD.101.074512}{\emph{Phys. Rev. D}
  {\bfseries 101} (2020) 074512}
  [\href{https://arxiv.org/abs/1908.06935}{{\ttfamily 1908.06935}}].

\bibitem{Ciavarella:2024fzw}
A.N.~Ciavarella and C.W.~Bauer,
  {}\href{https://doi.org/10.1103/PhysRevLett.133.111901}{\emph{Phys. Rev.
  Lett.} {\bfseries 133} (2024) 111901}
  [\href{https://arxiv.org/abs/2402.10265}{{\ttfamily 2402.10265}}].

\bibitem{Ciavarella:2025bsg}
A.N.~Ciavarella, I.M.~Burbano and C.W.~Bauer,
  {}\href{https://doi.org/10.1103/ylqb-phv5}{\emph{Phys. Rev. D} {\bfseries
  112} (2025) 054514} [\href{https://arxiv.org/abs/2503.11888}{{\ttfamily
  2503.11888}}].

\bibitem{Mueller:2024mmk}
N.~Mueller, T.~Wang, O.~Katz, Z.~Davoudi and M.~Cetina,
  {}\href{https://doi.org/10.1038/s41467-025-60177-7}{\emph{Nature Commun.}
  {\bfseries 16} (2025) 5492}
  [\href{https://arxiv.org/abs/2408.00069}{{\ttfamily 2408.00069}}].

\bibitem{Vodeb:2024tvo}
J.~Vodeb, J.-Y.~Desaules, A.~Hallam, A.~Rava, G.~Humar, D.~Willsch et~al.,
  {}\href{https://doi.org/10.1038/s41567-024-02765-w}{\emph{Nature Phys.}
  {\bfseries 21} (2025) 386}
  [\href{https://arxiv.org/abs/2406.14718}{{\ttfamily 2406.14718}}].

\bibitem{De:2024smi}
A.~De et~al., {} \href{https://arxiv.org/abs/2410.13815}{{\ttfamily
  2410.13815}}.

\bibitem{Cochran:2024rwe}
T.A.~Cochran et~al.,
  {}\href{https://doi.org/10.1038/s41586-025-08999-9}{\emph{Nature} {\bfseries
  642} (2025) 315} [\href{https://arxiv.org/abs/2409.17142}{{\ttfamily
  2409.17142}}].

\bibitem{Gonzalez-Cuadra:2024xul}
D.~Gonzalez-Cuadra et~al.,
  {}\href{https://doi.org/10.1038/s41586-025-09051-6}{\emph{Nature} {\bfseries
  642} (2025) 321} [\href{https://arxiv.org/abs/2410.16558}{{\ttfamily
  2410.16558}}].

\bibitem{Chen:2025zeh}
J.-W.~Chen, Y.-T.~Chen and G.~Meher, {}
  \href{https://arxiv.org/abs/2506.16829}{{\ttfamily 2506.16829}}.

\bibitem{Lee:2024jnt}
K.~Lee, F.~Turro and X.~Yao,
  {}\href{https://doi.org/10.1103/PhysRevD.111.054514}{\emph{Phys. Rev. D}
  {\bfseries 111} (2025) 054514}
  [\href{https://arxiv.org/abs/2409.13830}{{\ttfamily 2409.13830}}].

\bibitem{Farrell:2024fit}
R.C.~Farrell, M.~Illa, A.N.~Ciavarella and M.J.~Savage,
  {}\href{https://doi.org/10.1103/PhysRevD.109.114510}{\emph{Phys. Rev. D}
  {\bfseries 109} (2024) 114510}
  [\href{https://arxiv.org/abs/2401.08044}{{\ttfamily 2401.08044}}].

\bibitem{Davoudi:2024wyv}
Z.~Davoudi, C.-C.~Hsieh and S.V.~Kadam,
  {}\href{https://doi.org/10.22331/q-2024-11-11-1520}{\emph{Quantum} {\bfseries
  8} (2024) 1520} [\href{https://arxiv.org/abs/2402.00840}{{\ttfamily
  2402.00840}}].

\bibitem{Chai:2023qpq}
Y.~Chai, A.~Crippa, K.~Jansen, S.~K{\"u}hn, V.R.~Pascuzzi, F.~Tacchino et~al.,
  {}\href{https://doi.org/10.22331/q-2025-02-19-1638}{\emph{Quantum} {\bfseries
  9} (2025) 1638} [\href{https://arxiv.org/abs/2312.02272}{{\ttfamily
  2312.02272}}].

\bibitem{Farrell:2025nkx}
R.C.~Farrell, N.A.~Zemlevskiy, M.~Illa and J.~Preskill, {}
  \href{https://arxiv.org/abs/2505.03111}{{\ttfamily 2505.03111}}.

\bibitem{Schuhmacher:2025ehh}
J.~Schuhmacher, G.-X.~Su, J.J.~Osborne, A.~Gandon, J.C.~Halimeh and
  I.~Tavernelli, {} \href{https://arxiv.org/abs/2505.20387}{{\ttfamily
  2505.20387}}.

\bibitem{Davoudi:2025rdv}
Z.~Davoudi, C.-C.~Hsieh and S.V.~Kadam, {}
  \href{https://arxiv.org/abs/2505.20408}{{\ttfamily 2505.20408}}.

\bibitem{Farrell:2022vyh}
R.C.~Farrell, I.A.~Chernyshev, S.J.M.~Powell, N.A.~Zemlevskiy, M.~Illa and
  M.J.~Savage, {}\href{https://doi.org/10.1103/PhysRevD.107.054513}{\emph{Phys.
  Rev. D} {\bfseries 107} (2023) 054513}
  [\href{https://arxiv.org/abs/2209.10781}{{\ttfamily 2209.10781}}].

\bibitem{Chernyshev:2025lil}
I.A.~Chernyshev et~al.,
  {}\href{https://doi.org/10.1038/s41467-026-68536-8}{\emph{Nature Commun.}
  {\bfseries 17} (2026) 1826}
  [\href{https://arxiv.org/abs/2506.05757}{{\ttfamily 2506.05757}}].

\bibitem{Rosanowski:2025nck}
E.O.~Rosanowski, A.~Crippa, L.~Funcke, P.V.~Itaborai, K.~Jansen and S.~Singh,
  {} \href{https://arxiv.org/abs/2509.20558}{{\ttfamily 2509.20558}}.

\bibitem{Than:2024zaj}
A.T.~Than et~al.,
  {}\href{https://doi.org/10.1038/s41467-025-65198-w}{\emph{Nature Commun.}
  {\bfseries 16} (2025) 10288}
  [\href{https://arxiv.org/abs/2501.00579}{{\ttfamily 2501.00579}}].

\bibitem{Saner:2025nrq}
S.~Saner, O.~B{\u{a}}z{\u{a}}van, D.J.~Webb, G.~Araneda, C.J.~Ballance,
  R.~Srinivas et~al., {} \href{https://arxiv.org/abs/2507.19588}{{\ttfamily
  2507.19588}}.

\bibitem{Davoudi:2021ney}
Z.~Davoudi, N.M.~Linke and G.~Pagano,
  {}\href{https://doi.org/10.1103/PhysRevResearch.3.043072}{\emph{Phys. Rev.
  Res.} {\bfseries 3} (2021) 043072}
  [\href{https://arxiv.org/abs/2104.09346}{{\ttfamily 2104.09346}}].

\bibitem{Than:2025gso}
A.T.~Than, S.V.~Kadam, V.~Vikramaditya, N.H.~Nguyen, X.~Liu, Z.~Davoudi et~al.,
  {} \href{https://arxiv.org/abs/2509.11477}{{\ttfamily 2509.11477}}.

\bibitem{Crane:2024tlj}
E.~Crane et~al., {} \href{https://arxiv.org/abs/2409.03747}{{\ttfamily
  2409.03747}}.

\bibitem{Popov:2023xft}
P.P.~Popov, M.~Meth, M.~Lewenstein, P.~Hauke, M.~Ringbauer, E.~Zohar et~al.,
  {}\href{https://doi.org/10.1103/PhysRevResearch.6.013202}{\emph{Phys. Rev.
  Res.} {\bfseries 6} (2024) 013202}
  [\href{https://arxiv.org/abs/2307.15173}{{\ttfamily 2307.15173}}].

\bibitem{Illa:2024kmf}
M.~Illa, C.E.P.~Robin and M.J.~Savage,
  {}\href{https://doi.org/10.1103/PhysRevD.110.014507}{\emph{Phys. Rev. D}
  {\bfseries 110} (2024) 014507}
  [\href{https://arxiv.org/abs/2403.14537}{{\ttfamily 2403.14537}}].

\bibitem{Kurkcuoglu:2024cfv}
D.M.~K{\"u}rk{\c{c}}{\"u}oglu, H.~Lamm and A.~Maestri, {}
  \href{https://arxiv.org/abs/2410.16414}{{\ttfamily 2410.16414}}.

\bibitem{Gonzalez-Cuadra:2022hxt}
D.~Gonz{\'a}lez-Cuadra, T.V.~Zache, J.~Carrasco, B.~Kraus and P.~Zoller,
  {}\href{https://doi.org/10.1103/PhysRevLett.129.160501}{\emph{Phys. Rev.
  Lett.} {\bfseries 129} (2022) 160501}
  [\href{https://arxiv.org/abs/2203.15541}{{\ttfamily 2203.15541}}].

\bibitem{Meth:2023wzd}
M.~Meth et~al.,
  {}\href{https://doi.org/10.1038/s41567-025-02797-w}{\emph{Nature Phys.}
  {\bfseries 21} (2025) 570}
  [\href{https://arxiv.org/abs/2310.12110}{{\ttfamily 2310.12110}}].

\end{thebibliography}\endgroup

\end{document}